\documentclass[]{emulateapj}

\usepackage{rotating}

\usepackage{natbib}
\usepackage{latexsym}
\bibpunct{(}{)}{;}{a}{}{,}

\begin{document}

\title{An accreting millisecond pulsar with black hole-like X-ray variability: IGR~J00291+5934}

\author{Manuel Linares\altaffilmark{1}, Michiel van der Klis\altaffilmark{1}, Rudy Wijnands\altaffilmark{1}}

\date{}

\def\rem#1{{\bf (#1)}}
\def\hide#1{}

\altaffiltext{1}{Astronomical Institute ``Anton Pannekoek'', University of Amsterdam and Center for High-Energy Astrophysics, Kruislaan 403, NL-1098 SJ Amsterdam, Netherlands.}


\begin{abstract}
IGR~J00291+5934 is one of the seven accreting millisecond pulsars (AMPs) discovered so far. We report on the aperiodic timing and color analysis of its X-ray flux, using all the {\it RXTE} observations of the 2004 outburst. Flat-top noise and two harmonically related quasi-periodic oscillations, all of them at very low frequencies (0.01-0.1~Hz), were present in the power spectra during most of the outburst as well as a very high fractional variability ($\sim$50~\%). These properties are atypical not only for AMPs but also for neutron star low-mass X-ray binaries (LMXBs) in general. There are instead some remarkable similarities with the variability observed in black hole systems, reinforcing the connections between these two types of LMXB, as well as some interesting differences. We note finally that the results of this paper are difficult to reconcile with interpretations where any break frequency of power density spectra scales inversely with the mass of the central object at an accuracy sufficient to distinguish between the masses of neutron stars and black holes in LMXBs. 

\end{abstract}

\maketitle

\section{Introduction}
\label{sec:intro}

Most low-mass X-ray binaries (LMXBs) contain rapidly spinning neutron
stars (NSs), which only in a few cases become apparent as millisecond
X-ray pulsars. This growing sub-class of accreting millisecond pulsars
(AMPs; see \citealt{Wijnands05} for a review) consists of dim
transients that typically go in outburst during a few weeks once every
few years (so far the only exception is HETE~J1900.1-2455, still
active at the moment of writing, more than a year after its
discovery). AMPs are thought to be the evolutionary link between
neutron star LMXBs and millisecond radio pulsars, according to the
so-called recycling scenario \citep[][for a review]{BhatHeuv91}.
Eight years after the first of the AMPs was discovered
\citep[SAX~J1808.4--3658,][]{Wijnands98} it is still unknown why these
systems show pulsations and the rest of neutron star LMXBs do not,
although some explanations have been put forward (e.g. screening of
the magnetic field, \citealt{Cumming01}; smearing of the pulsations,
\citealt{Titarchuk02}).

Aperiodic variability in the X-ray lightcurves of black hole and
neutron star LMXBs (BH/NS-LMXBs), such as broadband noise or
quasi-periodic oscillations (QPOs), offers a powerful tool to study
the accretion flow in its inner region, where the bulk of the
radiation is thought to originate and where gravitational fields are
extreme \citep{vanderKlis00,vanderKlis06}. Using this tool NS-LMXBs
were classified according to their correlated timing and energy
spectral behavior into Z and atoll sources \citep{Hasinger89}. As
noted several years ago \citep[e.g.][]{Klis94}, very similar physical
mechanisms are expected to drive the accretion process in BH- and
NS-LMXBs, and the difficulties in finding unequivocal signatures of a
black hole or a neutron star in the energy or power spectral
properties of LMXBs are a natural consequence of this fact. Several
attempts to find direct BH signatures have been made based on the
spectral properties at high energies (\citealt{Sunyaev91}; but see
also \citealt{DiSalvo04} or \citealt{Barret00}) or in the power
spectral properties at high frequencies (\citealt{SunRev00}; but see
also \citealt{Kleinwolt04b}). On the other hand, apart from pulsations
and type I X-ray bursts, twin variable-frequency kilohertz
quasi-periodic oscillations (twin kHz QPOs) are now considered a NS
signature \citep{vanderKlis06}. NS-LMXBs are also known to be fainter
than BH-LMXBs at radio wavelengths \citep{Migliari06}, and BH-LMXBs
are generally fainter in X-rays when they are in quiescence \citep{Garcia01}.

We present in this work the results of the aperiodic timing and X-ray
color analysis of one of the seven known AMPs, IGR~J00291+5934. We
show the exceptional behavior of this source in its 2004 outburst,
more similar with respect to aperiodic X-ray variability to a BH-LMXB
than to the presently known NS-LMXBs: IGR~J00291+5934 was variable at
very low frequencies with a very high fractional amplitude. However,
previously identified NS signatures were also present in the highest
frequencies. In Section~\ref{sec:review} we briefly review some
information on this source and in Sections~\ref{sec:data} and
\ref{sec:results} we describe our analysis and give its results,
respectively. Section~\ref{sec:discussion} contains our final remarks
and conclusions.

\section{IGR~J00291+5934}
\label{sec:review}

On December $2^{nd}$, 2004, IGR J00291+5934 was discovered with {\it
INTEGRAL} during one of the Galactic plane scans
\citep{Eckert04}. Follow up observations with {\it RXTE} revealed
coherent pulsations at 598.88 Hz \citep{Markwardt04b}, making it the
fastest spinning AMP known. The pulse frequency exhibited a
sinusoidal modulation with a period of 147.4 minutes ($\sim$2.45 hr)
indicative of the orbital motion of the pulsar in the binary system
\citep{Markwardt04c}. Optical \citep{Fox04}, radio \citep{Pooley04,
Fender04, Rupen04} and infrared \citep{Steeghs04} counterparts were
identified and a recurrence time of $\sim$3 yr was suggested by
\citet{Remillard04}, based on {\it RXTE}'s all-sky monitor
data. Positive spin frequency derivatives were reported
by \citet{Falanga05b} and \citet{Burderi05}. Several estimates of the
distance to the source have been made, ranging from $\sim$3 to $\sim$10 kpc
\citep{Shaw05, Galloway05, Falanga05b, Jonker05, Burderi05}.

\section{Observations and Data Analysis}
\label{sec:data}

We used all the pointed PCA observations of the 2004 outburst of
IGR~J00291+5934, performed by {\it RXTE} between December 3d. and 20th
(see Table~\ref{table:obs}). An intermediate
polar\footnote{Intermediate polars \citep[][for a review]{Patterson}
are a sub-class of accreting white dwarfs with a magnetic field strong
enough to channel the accretion flow onto the magnetic poles (thereby
emitting X-rays pulsed at the spin period) but unable to lock the
rotation of the white dwarf with the orbital motion (as is the case
for the so-called polars). Chance placed two semidetached binaries
emitting X-ray pulses inside the PCA field of view, one of them
containing a white dwarf and the other a neutron star.} was also
inside the PCA's field of view: V709 Cas (which has a spin period of
$\sim$313 seconds; \citealt{Haberl95,
Norton99,Falanga05c}). Hereinafter we refer to this system as the
IP. All necessary caveats were taken in both color and timing analysis
in order to avoid contamination from this foreground source, as
explained below.

\begin{figure}[b]

  \resizebox{0.7\columnwidth}{!}{\rotatebox{0}{\includegraphics[]{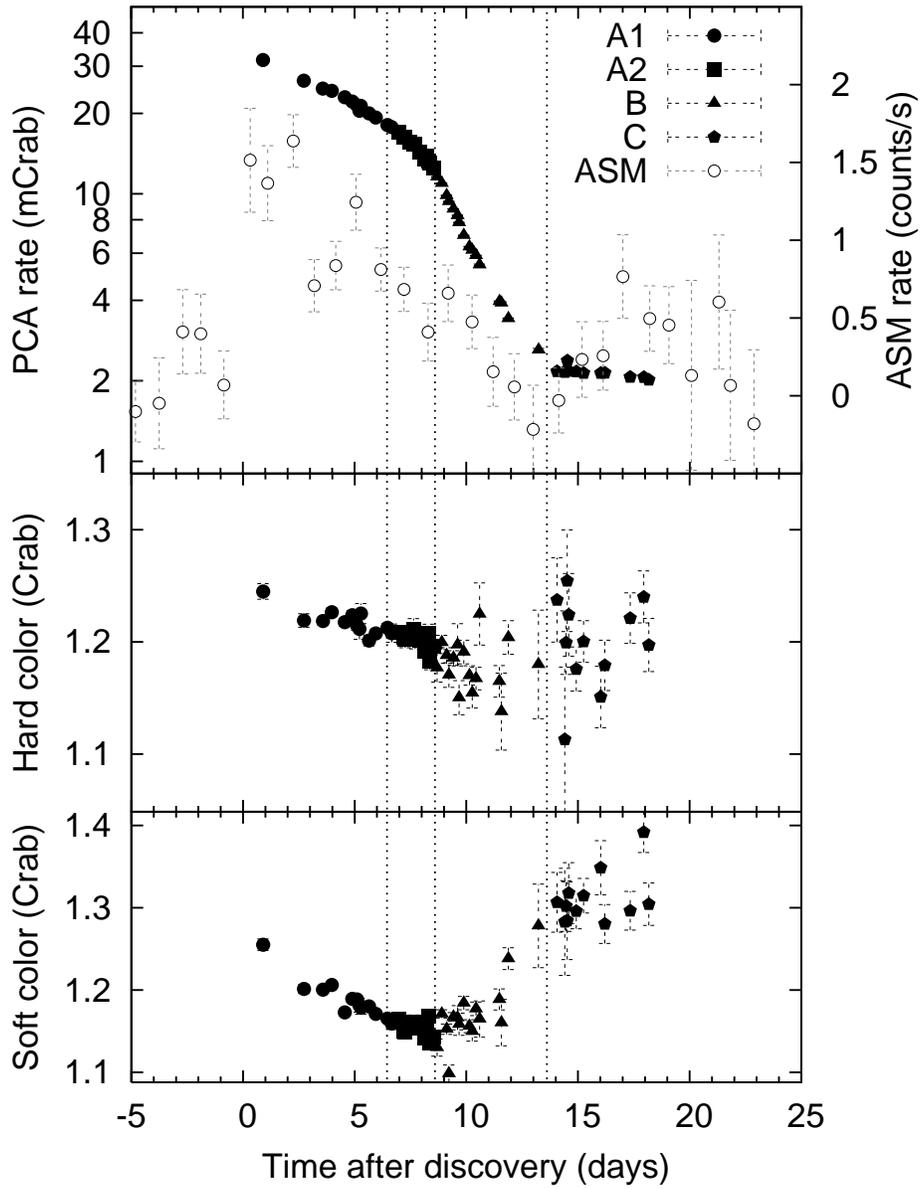}}}

  \caption{Time evolution of colors and intensity during IGR~J00291+5934's December 2004 outburst. The upper panel shows the lightcurve from both ASM (1.5-12 keV, daily average; open circles) and pointed PCA (2-16 keV, observation average; filled symbols) data. Vertical dotted lines show the borders between averaged data sets A1, A2, B and C. The increase in colors after day $\sim$10 is likely due to foreground contamination (see text).}
    \label{fig:lc}
\end{figure}

After filtering out unsuitable data according to the recommended
criteria\footnote{Elevation from the Earth greater than 10 degrees and
pointing offset lower than 0.02 degrees; see PCA digest at {\it
http://heasarc.nasa.gov/docs/xte/pca\_news.html} or, e.g.,
\citet{Straaten03}} we extracted count rates from Standard 2 data,
with 16 s time resolution, in the following bands (in keV): A:
2.0--3.5, B: 3.5--6.0, C: 6.0--9.7, D: 9.7--16.0, and we computed soft
(B/A) and hard (D/C) colors and intensity (A+B+C+D). The instrumental
background was subtracted according to the standard faint-source
background models and the colors and intensity normalized to the Crab
values nearest in time \citep{Kuulkers94} and in the same PCA gain
epoch \citep[e.g.,][]{Straaten03}. Using WebPimms and the spectrum
reported by \citet{Falanga05c}, and taking into account the $\sim$0.3
deg. offset between the IP and our source, we estimate a foreground
2-16 keV count rate from the IP of $\sim$3.6 c/s/PCU (to be
compared with the peak rate in our observations of $\sim$76 c/s/PCU ),
which matches approximately the observed intensity during our last
dataset ($\sim$5 c/s in PCU2), when IGR~J00291+5934 had presumably ceased
activity (see Section~\ref{sec:results}, Figure~\ref{fig:lc}). The
effect of this contaminating flux is probably strongest on the colors,
which are sensitive to subtle changes in the energy spectrum. For that
reason we consider only the color-color diagram of the first half of
the outburst, where the flux from the IP is less than 20\% of the
background subtracted total (see Section~\ref{ssec:color}).

GoodXenon and Event modes were used for the timing analysis (see
Table~\ref{table:obs}), rebinned in both cases to a 1/8192-s time
resolution (Nyquist frequency of 4096~Hz) and including only the
counts in the $\sim$2.5-30 keV range (absolute channels 5-71; this
choice optimizes the S/N). We performed fast Fourier transforms in
1024-s data segments, reaching thereby Fourier frequencies down to
$\sim$0.001~Hz.  Two observations were shorter than 1024 s
(90425-01-01-01 and 90425-01-01-03) so we performed and analyzed a
512-s FFT in each of them. No background subtraction or dead-time
correction were made prior to the FFTs and the Poisson noise power was
subtracted from the resulting power density spectra, following
\citet{Kleinwolt04}: we first estimated the Poisson noise using the
expression proposed by \citet{Zhang95} and then (after inspecting the
$\sim$2--4~kHz range and finding no unexpected features) shifted it to
match the level between 2000--4096~Hz, where no intrinsic power is
expected to be present, but only counting statistics noise (this shift
was in all cases smaller than 0.06\% of the previously estimated
Poisson level). We then normalized the power spectra in the so-called
rms normalization \citep{vanderklis95b}. For this normalization
purpose, in order to take into account the foreground flux contributed
by the IP, we used the count rates in the tail of the outburst (when
the AMP reaches its quiescent state; set C; see Fig.~\ref{fig:lc}) to
estimate the average background and foreground rate during the data
segments used for timing. In order to improve the statistics we
averaged observations successive in time and with power spectra
consistent within errors, thereby dividing the outburst data into four
sets (Table~\ref{table:obs}, Fig~\ref{fig:lc}). The integrated power
was calculated between 0.01 and 100 Hz in order to avoid contamination
from the IP's variability (see Section~\ref{ssec:timing}).

We fitted the power spectra with a fit function composed of a sum of
Lorentzians in the ``$\nu_{max}$ representation''\footnote{In this
representation \citep{Belloni02}, if $\nu_0$ is the Lorentzian's
centroid frequency and $\Delta$ its HWHM (half width at half maximum),
$\nu_{max}=\sqrt{\nu_0^2+\Delta^2}$ gives the characteristic frequency
of the feature (near the centroid if it is narrow and near the
half-width if it is wide). The quality factor $Q=\nu_0/2\Delta$ is
used in this representation as a measure of the coherence of the
variability feature. Its strength is given by the integral power
(0--$\infty$) whose square root, in the normalization we use, is the
fractional rms amplitude of the variability.}, excluding the first
four frequency bins to preclude the IP's variability. In some cases,
in order to avoid a meaningless negative coherence, $Q$ was fixed to
zero, which is equivalent to fitting a zero-centered Lorentzian. From
now on, following previous work \citep[][]{Belloni02, Straaten03,
vanderKlis06, Altamirano05, Linares05} we refer to these components as
$L_i$, where the L stands for Lorentzian and '$i$' is the label
identifying the component (such identification is given in
Sections~\ref{ssec:timing} and \ref{ssec:NSBH}). Following this
notation we call $L_i$'s characteristic frequency $\nu_i$ and its
coherence $Q_i$. Five to seven Lorentzians were necessary to fit the
power spectra, as described in detail in the next section.

\section{Results}
\label{sec:results}

\subsection{Colors and intensity}
\label{ssec:color}

The time evolution of the count rate and colors can be seen in
Fig.~\ref{fig:lc}. The lightcurve shows two clearly different slopes
in a semi-logaritmic plot, indicating two different exponential decay
timescales. As noted in previous works \citep[e.g.][]{Wijnands05},
this is common behavior among AMPs, with XTE~J1807--294
the only known exception \citep{Linares05, Falanga05}. We fitted the
first two parts of the (2.0--16.0 keV) lightcurve (A1+A2 and B) and
found e-folding times of $7.9\pm0.3$ and $2.45\pm0.05$ days, similar to
the ones reported in \citet{Falanga05b} from the {\it INTEGRAL}
20--100 keV lightcurve.

\begin{figure}[b]
  \resizebox{0.7\columnwidth}{!}{\rotatebox{-90}{\includegraphics[]{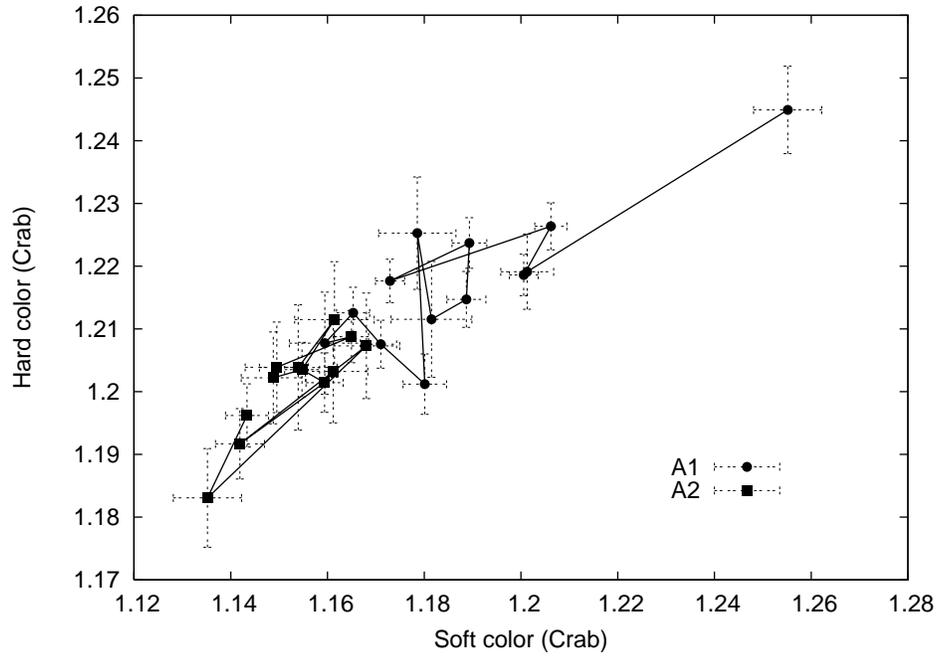}}}

  \caption{Color-color diagram of data sets A1 and A2, where the colors are not yet dominated by the intermediate polar in the field of view. The upper right point corresponds to the first pointed observation and the line connects observations successive in time.}
    \label{fig:ccd}
\end{figure}

Overall the source spectrum was hard compared to that of other AMPs
\citep{Straaten05, Linares05} and it softened at both high (6.0--16.0
keV) and low (2.0--6.0 keV) energies during the first $\sim$9 days
decay (see time evolution of soft and hard color in
Fig.~\ref{fig:lc}).  The resulting track in the color-color diagram in
this first half of the outburst is shown in
Figure~\ref{fig:ccd}. After the first 9 days (sets B and C in our
analysis), the IP contributes with a significant fraction of the flux
($\gtrsim$20\% if all the flux in C comes from the IP) and the
increase of the colors is likely due to that \citep[intermediate
polars are thought to be the hardest X-ray emitting cataclysmic
variables; see e.g.][]{Martino04}.

\subsection{Aperiodic timing}
\label{ssec:timing}

\begin{figure}[b]
  \resizebox{0.5\columnwidth}{!}{\rotatebox{-90}{\includegraphics[]{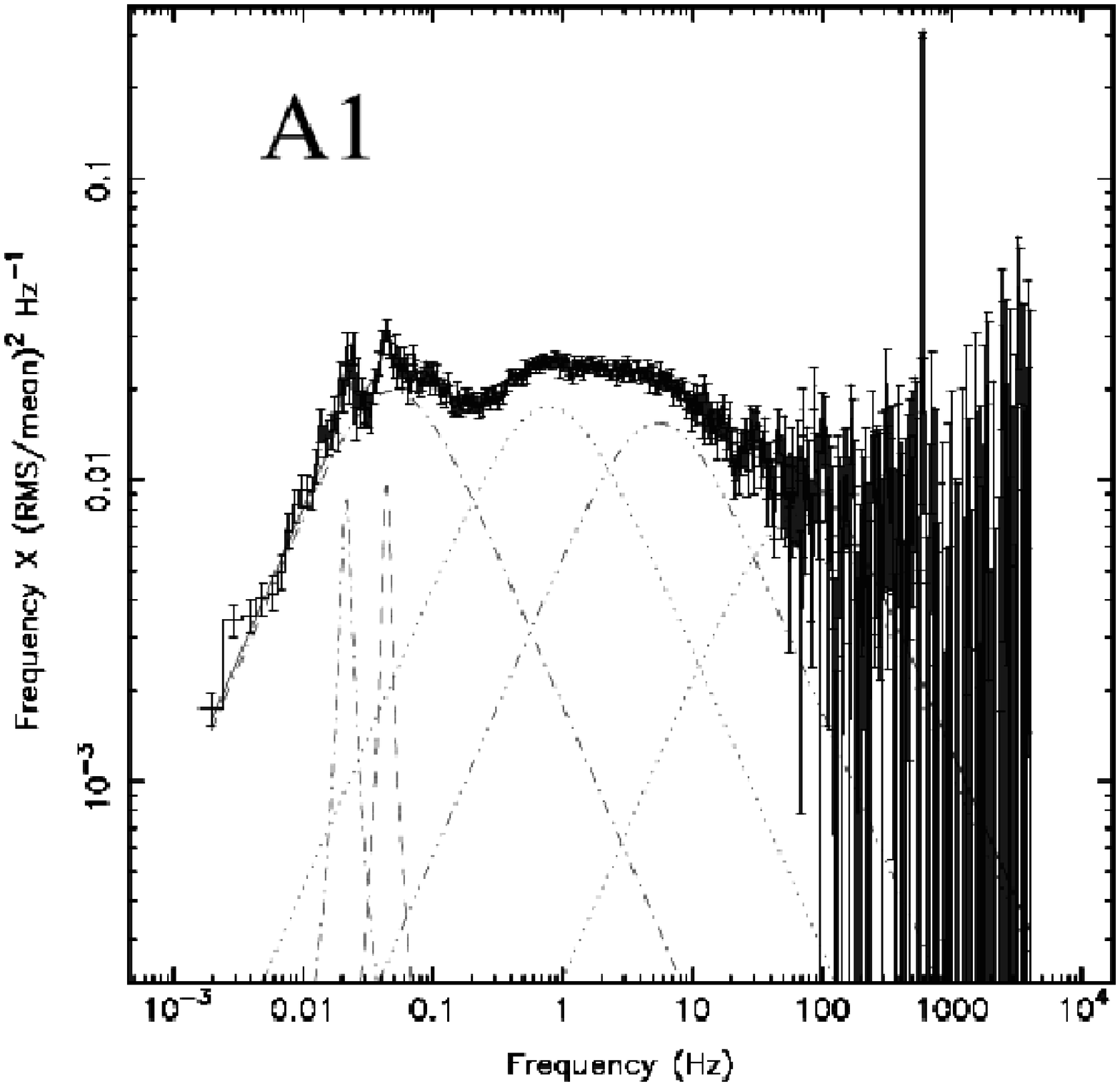}}}
  \resizebox{0.5\columnwidth}{!}{\rotatebox{-90}{\includegraphics[]{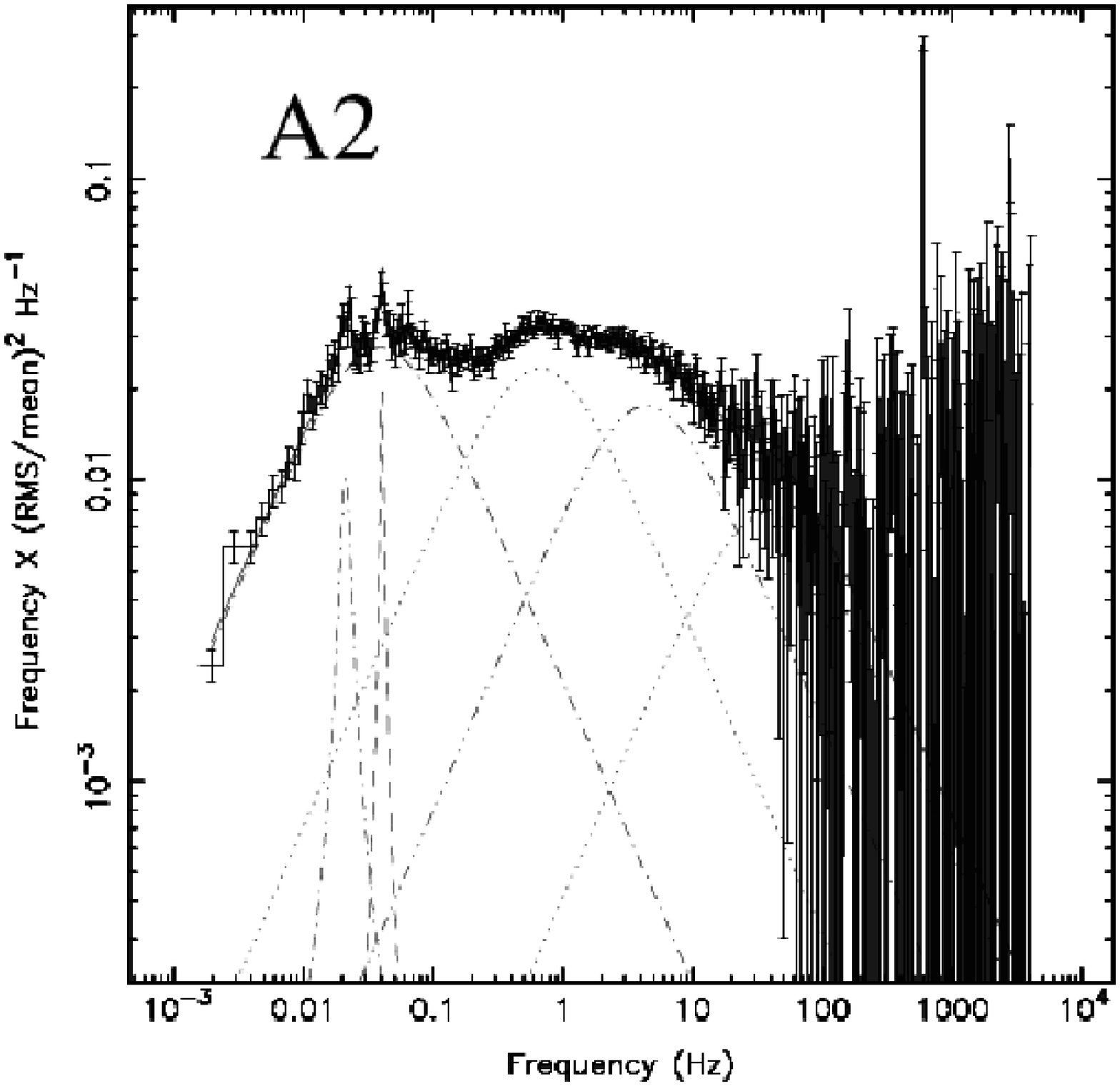}}}
  \resizebox{0.5\columnwidth}{!}{\rotatebox{-90}{\includegraphics[]{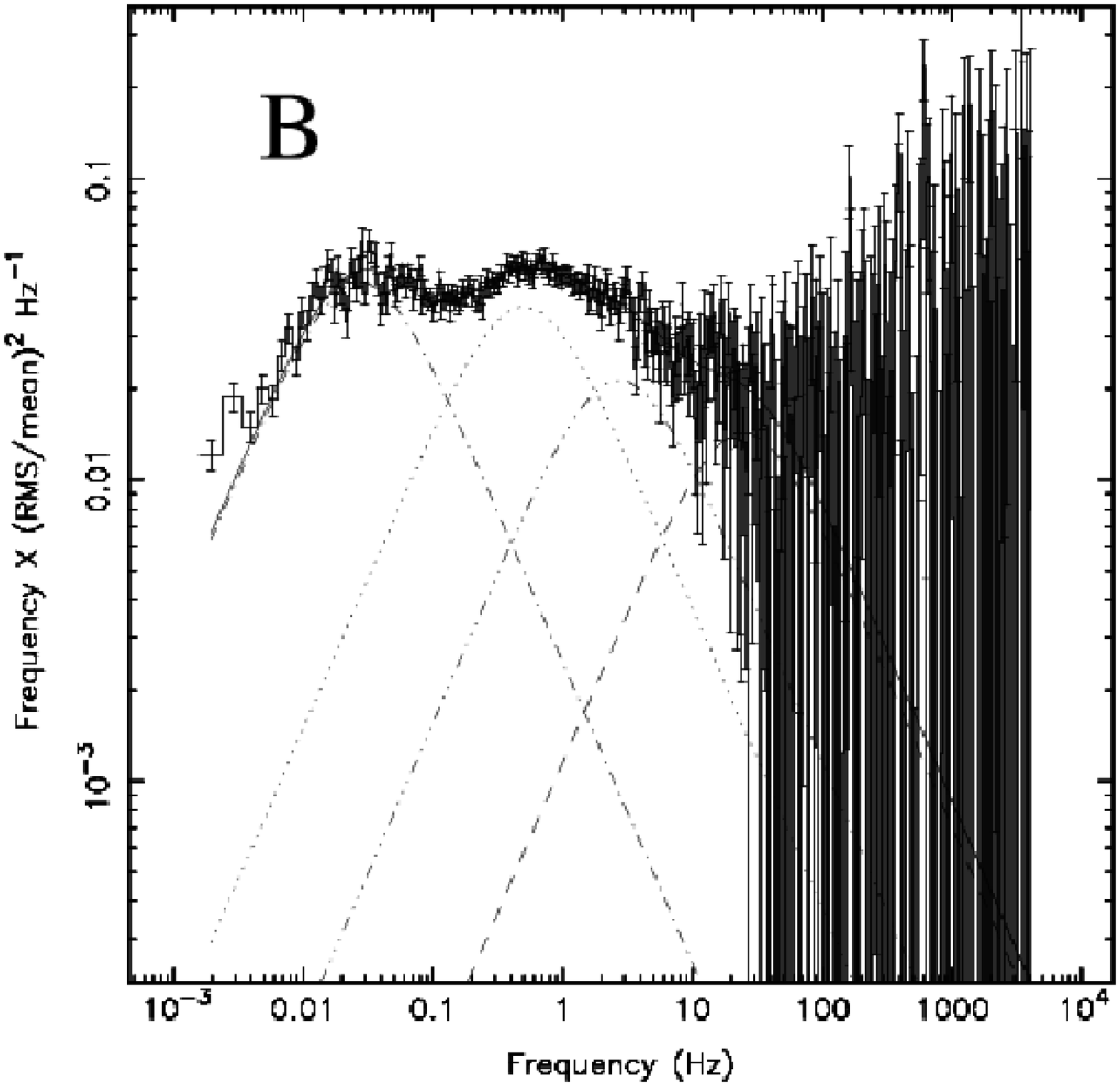}}}
  \caption{Power spectra in Power$\times$Frequency vs. Frequency representation. Also shown are the respective fit functions and their Lorentzian components. The spin frequency of the neutron star shows as a spike at $\sim$600Hz.}
    \label{fig:ps1}
\end{figure}

\begin{figure}[b]

  \resizebox{0.7\columnwidth}{!}{\rotatebox{0}{\includegraphics[]{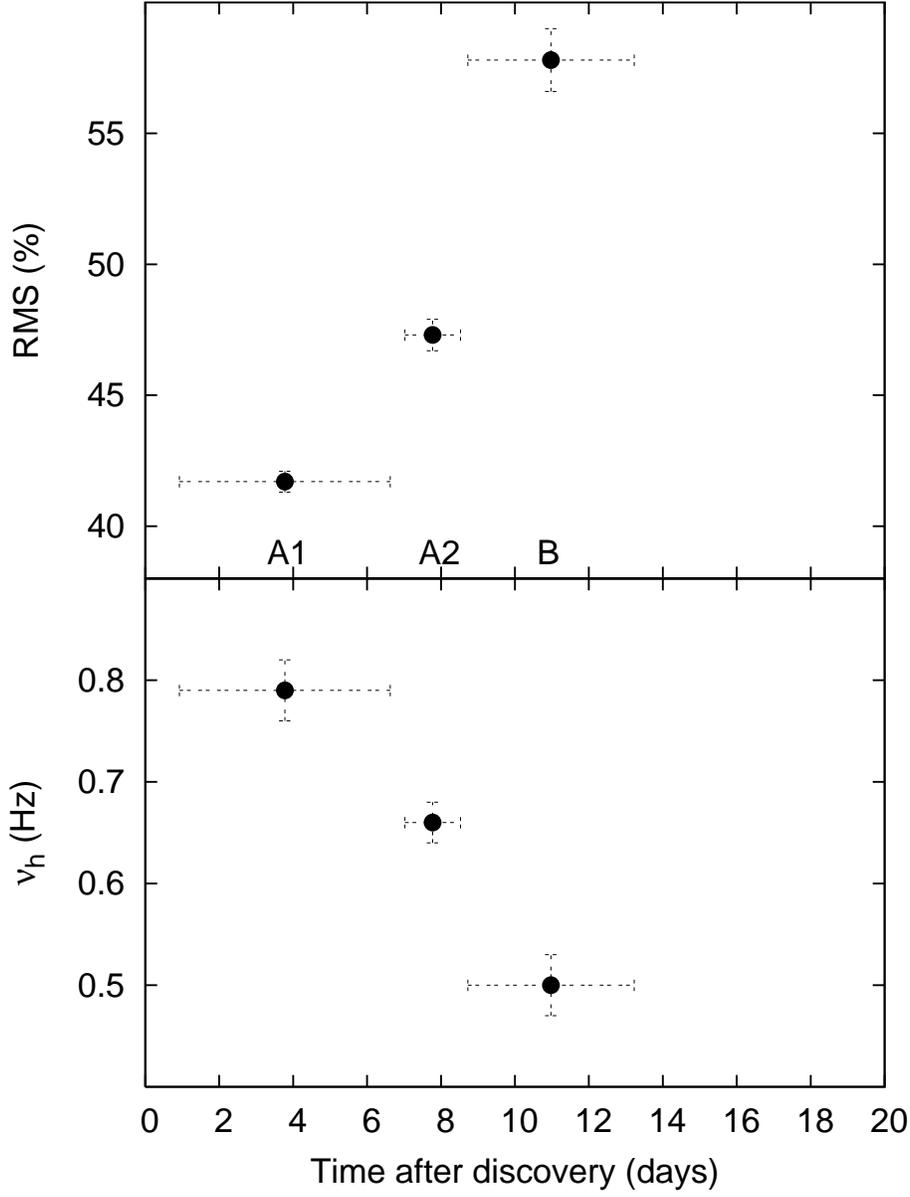}}}

  \caption{Integrated (0.01--100 Hz) fractional rms amplitude and characteristic frequency of the ``hump'' in the power spectra ($L_h$) for the three first datasets. Error bars in the X axis indicate the time interval of each dataset.}
    \label{fig:rmsqpo}
\end{figure}

The power spectra (Fig.~\ref{fig:ps1}) on first inspection show an
overall shape characteristic of low (luminosity) and hard (spectra)
states of atoll sources, usually called island and extreme island
states (where spectral hardness and total rms increase and luminosity
and variability frequencies decrease when going from the island to the
extreme island state). However, the overall (0.01-100 Hz) fractional
rms amplitude is in the $\sim$42-58 \% range (Fig.~\ref{fig:rmsqpo},
Table~\ref{table:psfits}), much higher than any value reported to date
for a NS-LMXB (see e.g. \citealt{Olive98, Barret00, Belloni02}). Five
to six Lorentzian components were used to fit these power spectra. The
best fit parameters are displayed in Table~\ref{table:psfits}. The fit
functions comprise i) one broad Lorentzian at 0.03-0.05~Hz accounting
for the flat-top noise and its break ($L_b$ for ``break''; see
Sec.~\ref{sec:data} for nomenclature) ii) two QPOs between 0.02 and
0.05~Hz ($L_{QPO}$ and its subharmonic, $L_{QPO/2}$) and iii) three
zero-centered Lorentzians with characteristic frequencies covering the
0.5-0.8 Hz, 3-5.5 Hz and 25-70 Hz ranges (possibly $L_h$, $L_{\ell
ow}$ and $L_u$, respectively; see Section~\ref{ssec:NSBH} for further
discussion). All components are significant at a $>3\sigma$ level
except $L_{QPO}$ and $L_{QPO/2}$ in dataset B.

\begin{figure}[b]
  \begin{center}

  \resizebox{0.6\columnwidth}{!}{\rotatebox{0}{\includegraphics[]{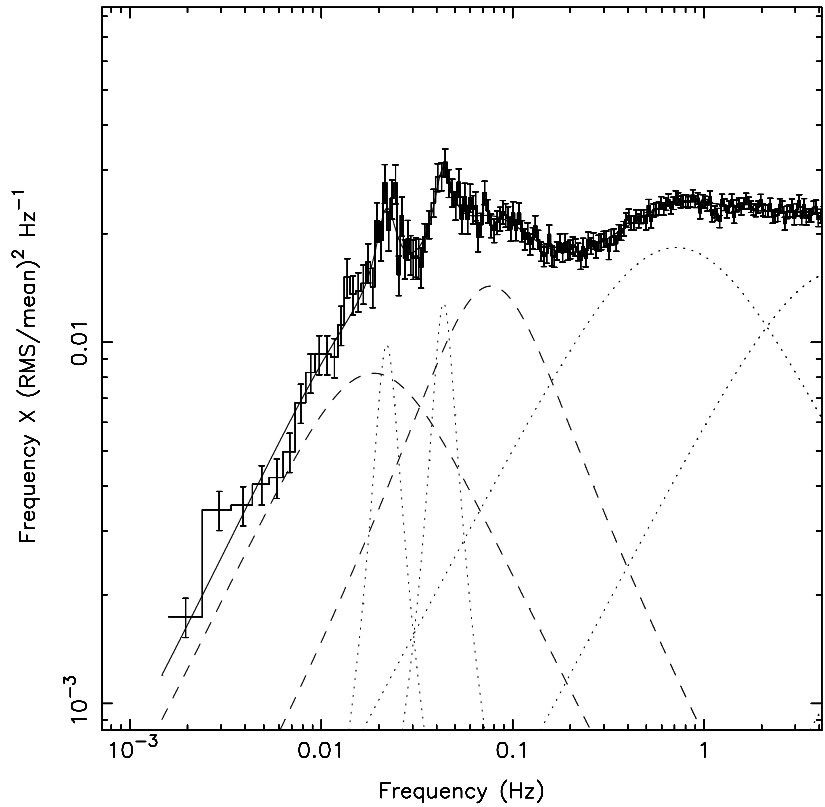}}}

  \caption{Zoom into the low-frequency part of the power spectrum of set A1, showing an alternative function which uses two broad lorentzians (dashed lines) to fit the break. See Section~\ref{sec:data} for details.}
    \label{fig:psalt}
 \end{center}
\end{figure}

The flat-top noise present in these observations can be alternatively
described with a sum of two broad Lorentzians, one of them giving the
first break and the other describing a further steepening above that
break (a ``wing'' above the break in Frequency$\times$Power
representation; see Figure~\ref{fig:psalt}). For completeness we
report the parameters of those fits in
Table~\ref{table:psfits2}. These fits are better from a statistical
point of view ($4.7\sigma$, $2.3\sigma$ and $3\sigma$ improvement in
set A1, A2 and B, respectively, according to the F-test) and give a
more detailed description of the structure around the break, but are
more difficult to compare with other sources (where usually
single-Lorentzian or broken power-law breaks are measured). Therefore,
we use the frequencies of the first described six-Lorentzians fit
function when comparing with frequencies of other sources (see
Section~\ref{sec:discussion}).

\begin{figure}[b]

  \resizebox{0.5\columnwidth}{!}{\rotatebox{-90}{\includegraphics[]{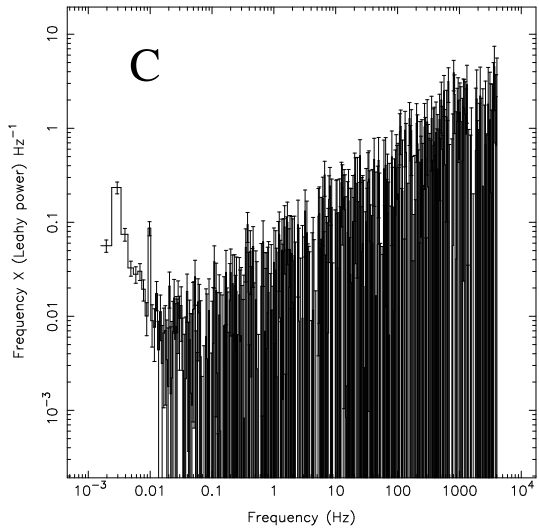}}}

  \caption{Power spectra of the background/foreground dominated set, in Power$\times$Frequency vs. Frequency representation. The spin frequency of the white dwarf in the field of view is visible at $\sim$0.003Hz (312.8s), as well as its second harmonic. Both periodicities have been already detected in this system in the 0.1-2.4 keV {\it ROSAT} energy band \citep{Haberl95, Norton99}}
    \label{fig:ps2}
\end{figure}

Assuming the IP did not drastically vary its behavior synchronously to
the AMP's outburst phases, we can reject the possibility that an
appreciable part of the variability above 0.01~Hz comes from the
IP. This is shown qualitatively by the lack of power above
$\sim$0.01~Hz in the power spectrum of the last dataset
(Figure~\ref{fig:ps2}), when the AMP is back to
quiescence. Quantitatively, if e.g., $L_{QPO}$ came from the IP it had
an rms amplitude of $\sim$27\% of the IP's flux in set A1, and fell
below the detection level in set C. A similar quantitative argument
can be made for all variability components, as the integrated power
between 0.01 and 100~Hz in set C was zero within errors. We split the
energy range into four contiguous bands (channels 5-9, 10-15, 16-25,
26-71: 2.5--4 keV, 4.5--6.5 keV, 6.9--10.6 keV and 11.0--30.0 keV,
respectively) and found no significant variations with energy in the
strength, width or frequency of the detected QPOs or noise
components. On the other hand a clear time evolution of both the QPOs
and the broad-band noise is detected as shown in Fig.~\ref{fig:rmsqpo}:
the (0.01-100~Hz) fractional rms variability increased during the
outburst (from 41.7 to 57.8\%), while the frequencies of the QPOs and
the other components decreased (from $\sim$0.79 to $\sim$0.50~Hz for
$L_{h}$).

\section{Discussion}
\label{sec:discussion}

\subsection{Exceptional island state}
\label{ssec:extreme}

Our analysis of the fastest spinning AMP, IGR~J00291+5934, reveals the
exceptional behavior of this source in its 2004 outburst compared to
the rest of NS-LMXBs: very strong X-ray variability at very low
Fourier frequencies. Its power spectra showed flat-topped noise
(i.e. approximately flat below a certain ``break'' frequency and
decreasing above that) with extremely low break frequencies
($\sim$0.04~Hz), as well as very high integrated fractional rms
variability ($\sim$50~\%). These break frequencies are about two
orders of magnitude lower than those observed in Z sources and about
one order of magnitude lower than those previously found in extreme
island (low-hard) states of atoll sources
\citep[e.g,][]{Barret00,Belloni02}\footnote{A reanalysis of
GS~1826--238 data in \citet{Barret00} led to a break frequency above
0.1~Hz}. The measured fractional rms amplitude of the variability is
the highest found so far for in a NS-LMXB. Both the break frequencies
and fractional variability observed in IGR~J00291+5934 are instead
similar to those detected in low-hard states of BH-LMXBs (though the
frequencies are low even for the lowest BH-LMXB range; see next
section for a more detailed comparison). It has been noted
\citep{Straaten05, Linares05} that AMPs hardly ``make it to the soft
state'' and instead remain in (low-hard) island or extreme island
states. The (2.5--25.0~keV) peak luminosity of IGR~J00291+5934
(between 0.5 and 5.5~\% of the Eddington luminosity according to the
flux measured by \citealt{Galloway05} and the different distance
estimates, see Sec.~\ref{sec:review}) was similar to that of other
transient NS-LMXBs when they show ``canonical'' extreme island states,
which seems to indicate that the exceptional state of our source was
not purely a consequence of an exceptionally low accretion rate. 

The lack of high frequency variability ($\gtrsim$100~Hz) during the
outburst of IGR~J00291+5934 might be related to a combination of a
dynamically important magnetic field (needed in AMPs in order to
produce the observed pulsations) and a fast spin (spin frequency of
$\sim$600 Hz in the case of IGR~J00291+5934) that could prevent the
accretion disk from reaching the innermost regions, where the highest
dynamical frequencies are present. This could in turn be related to
the strong overall variability that we measured. If, as has been
argued \citep[e.g.][]{Churazov01, Gilfanov03}, the disk sets the
variability frequencies and the comptonizing medium (corona, boundary
layer, spreading layer, ADAF or similar) sets the variability
amplitude, then a disk truncated at a large radius could provide both
low dynamical frequencies and strong {\it fractional} variability, as
its flux contribution would be relatively low and hence so would be
the decrease that this flux would cause in the relative amplitude of
the variability. We note, however, that in standard descriptions
\citep[e.g.][]{Gosh78} the inner disk radius does not depend on the
spin rate, and if the so-called propeller mechanism is responsible for
clearing out the inner disk annulus this would be in contradiction
with the reported spin-up \citep{Falanga05b, Burderi05}, as the
expelled matter would remove angular momentum from the neutron
star. The interaction between a rapidly spinning magnetized neutron
star and an accretion disk is a complex problem.  Most of the work so
far concerns the ``classical'' slow X-ray pulsars \citep{Lamb73,
Gosh78} and only recently the AMP case has received more attention
\citep{Psaltis99c,Rappaport04}.

\begin{figure}[b]
  \resizebox{0.8\columnwidth}{!}{\rotatebox{-90}{\includegraphics[]{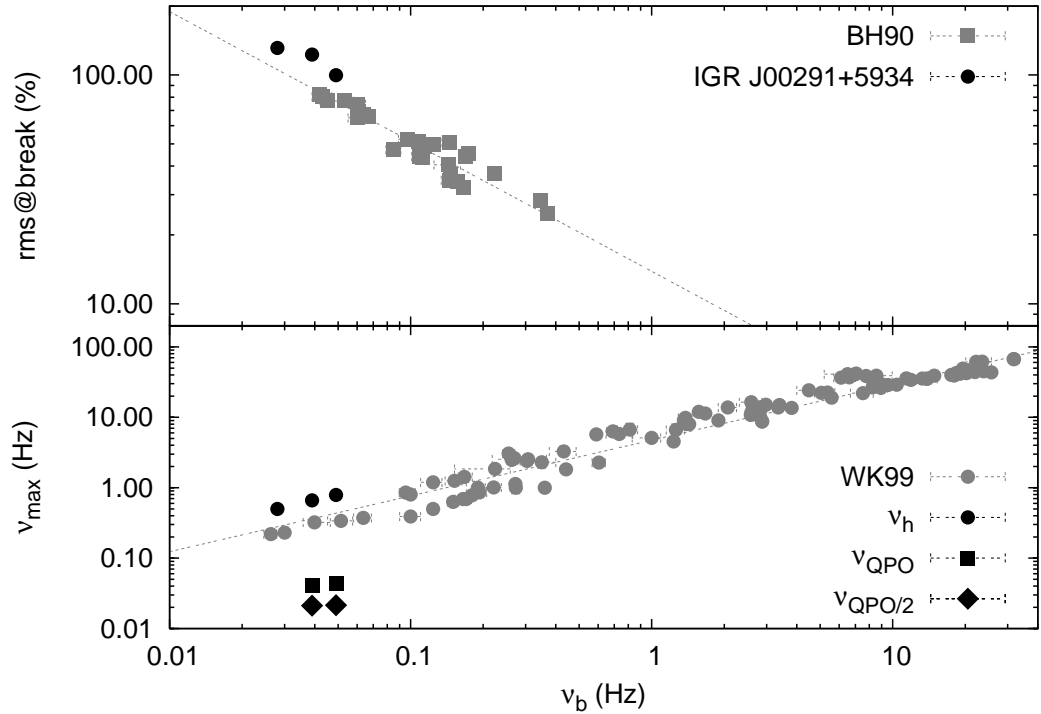}}}

  \caption{{\it (Top:)} Correlation between rms level at the break of the power spectra and the frequency of this break \citep[after][BH90: Cyg X-1, grey squares]{BH90}. {\it (Bottom:)} Correlation between the frequencies of $L_b$ and $L_h$, after \citet{WK99}. The lines indicate the best power law fits to the original correlations and the points that come from our work are shown in black.}
    \label{fig:BHWK}
\end{figure}

 An anticorrelation in the flat-topped noise of Cyg~X-1 was discovered
by \citet{BH90} linking the rms level of the flat top with the break
frequency: the lower the break frequency, the higher is the flat-top
(in rms normalization). In Figure ~\ref{fig:BHWK} (top) we see how
IGR~J00291+5934 extends this relation to the lowest frequencies and
the highest rms levels, even when comparing with the BH system Cyg~X-1
\citep[see][for an update of this correlation and comparison with
other neutron star systems, which are all at frequencies $>$0.1~Hz,
and the BH-LMXB XTE~J1118+480, which is contiguous with but still at
slightly higher frequecies than our NS-LMXB
IGR~J00291+5934]{Belloni02}.

\subsection{IGR~J00291+5934 vs. BH-LMXBs}
\label{ssec:NSBH}

\begin{figure}[b]

  \resizebox{0.32\columnwidth}{!}{\rotatebox{0}{\includegraphics[]{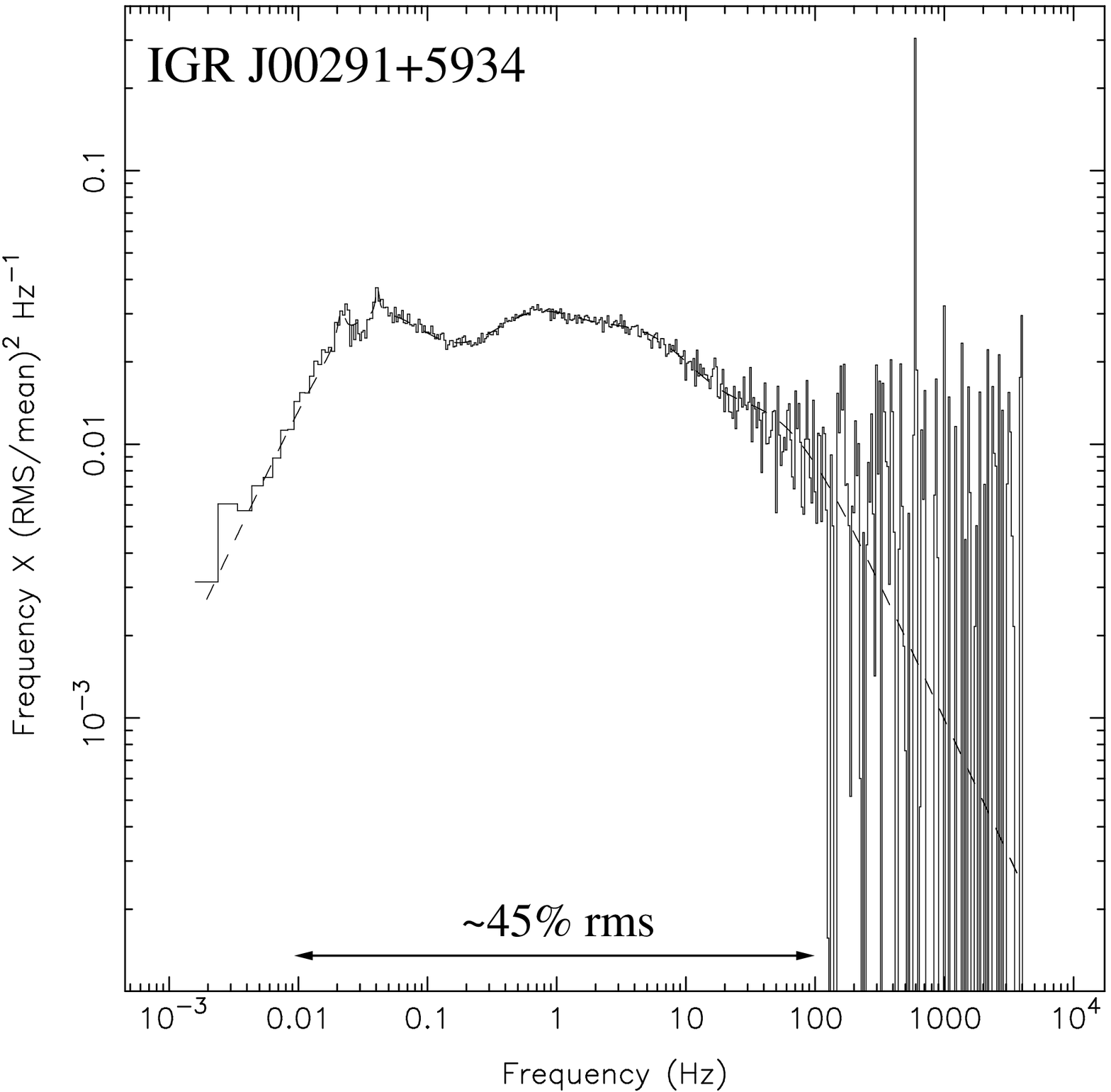}}}
  \resizebox{0.32\columnwidth}{!}{\rotatebox{0}{\includegraphics[]{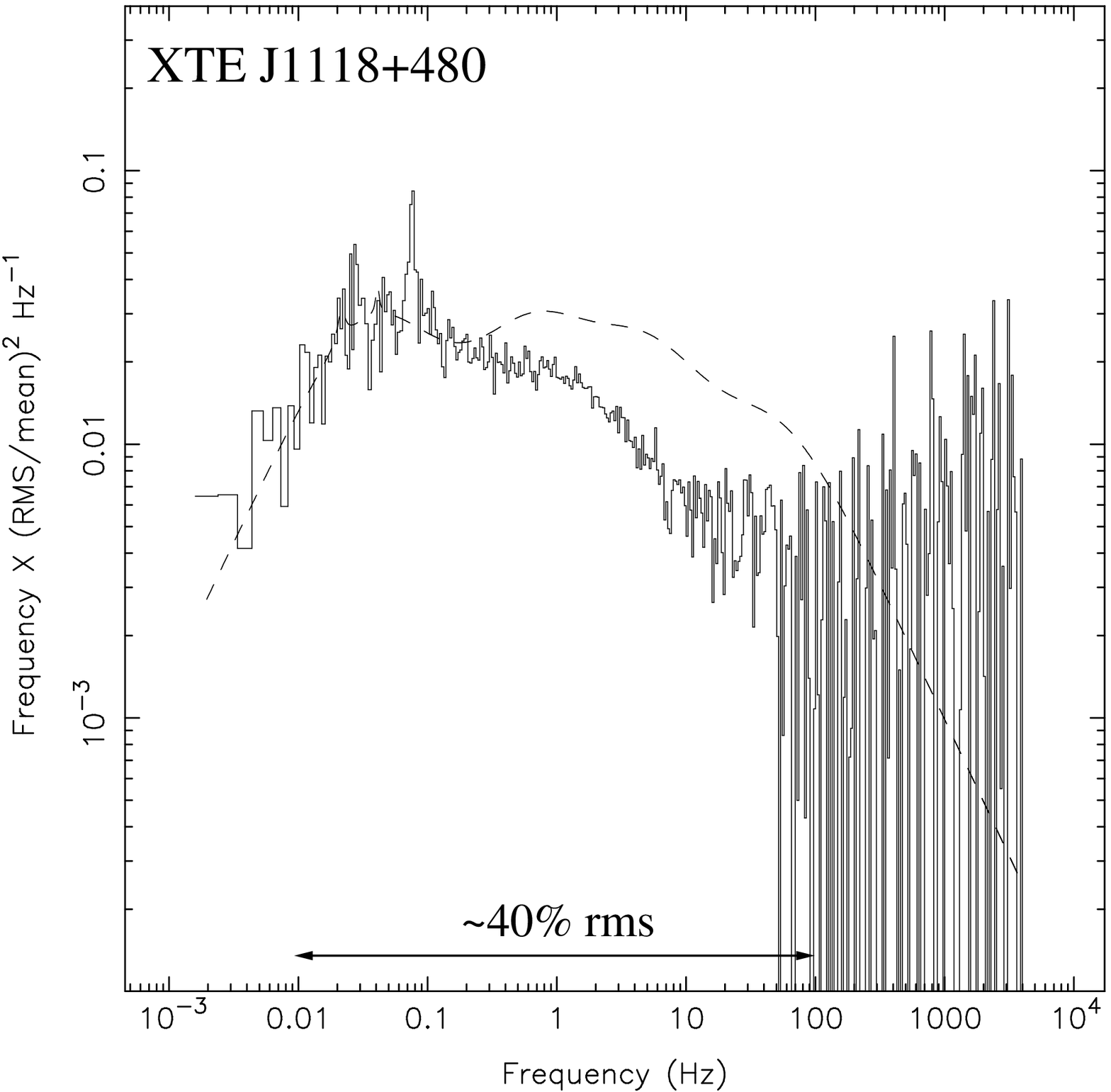}}}
  \resizebox{0.32\columnwidth}{!}{\rotatebox{0}{\includegraphics[]{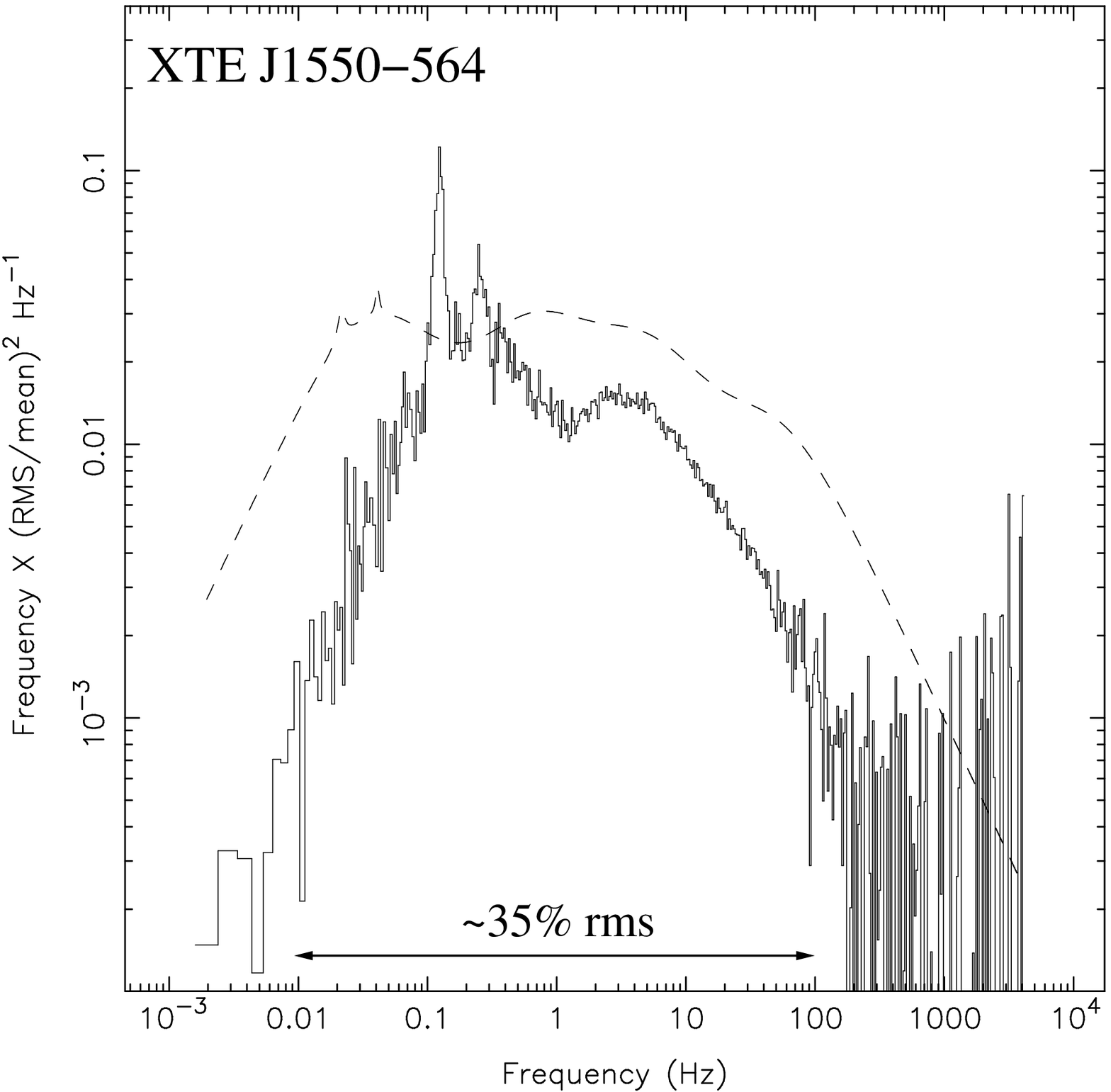}}}

  \caption{Power spectra of the accreting millisecond pulsar IGR~J00291+5934 and of two black hole low-mass X-ray binaries in the low state (XTE J1118+480, \citealt{Revni00}; XTE J1550-564, \citealt{Cui99}). The power spectrum of IGR~J00291+5934 corresponds to sets A1, A2 and B of our analysis. The energy band of XTE J1550-564's power spectrum ($\sim$2--13 keV) is different than the one used in the other two cases ($\sim$2.5--30 keV). The fractional rms variability (0.01-100 Hz) is shown on each plot. The mimetism between the neutron star and the black hole candidates is obvious, although there are also interesting differences (see Section~\ref{ssec:NSBH}). }
    \label{fig:comp}
\end{figure}

The similarities between IGR~J00291+5934 and BH-LMXBs in low-hard
states are illustrated in Figure~\ref{fig:comp}, where a power
spectrum of IGR~J00291+5934 is compared to one of XTE~J1118+480 and
one of XTE~J1550-564 \citep[e.g.][respectively]{Revni00, Cui99}. The
flat-top noise has very similar power, break frequency and QPOs. There
are also obvious yet interesting differences: as previously noted in a
study of low states of BH/NS-LMXBs the NS power spectra show more
power at high frequencies ($\gtrsim$10~Hz) than the BH ones
\citep{SunRev00}. In this extreme case, however, the differences
between the power spectra of IGR~J00291+5934 and XTE~J1118+480 start
at $\sim$0.2 Hz, where the NS power spectrum rises and the BH one
starts to fall (see Fig.~\ref{fig:comp}). The power spectrum of
XTE~J1550-564 has a very similar shape to that of IGR~J00291+5934 and
XTE~J1118+480, but shifted towards higher frequency and lower rms. It
also shows a prominent bump at $\sim$1-10~Hz (similar to what is seen
in IGR~J00291+5934 at slightly lower frequency) followed by a steep
decline above $\sim$10~Hz (characteristic of BH systems; see
\citealt{Kleinwolt06} for an extensive study and further
discussion). In contrast to the ones studied in \citet{SunRev00} this
NS-LMXB shows no significant power above $\sim$100~Hz.

We found two low-frequency QPOs at $\sim$0.02 and $\sim$0.04~Hz,
similar to the ones seen in BH- and other NS-LMXBs in low-luminosity
states \citep{Casella05, Straaten05}. Their centroid frequency ratios
were 2.06$\pm$0.06 and 1.91$\pm$0.06 in A1 and A2 respectively,
indicating an harmonic relation. In general this kind of QPO is
associated with low-luminosity states and therefore it requires a
relatively low mass accretion rate in the inner regions of the
disk. Several attemps have been made to investigate the nature of
these low-frequency QPOs, e.g. in the context of shot-noise models
\citep{Vikh94} or thermal-viscous instabilities \citep{Chen94}. In any
case our work shows clearly that neither the presence of a solid
surface nor a magnetic field affects this phenomenon so that the same
physical mechanism must be at work in NS and BH systems in order to
produce this common feature: harmonically related QPOs superposed on a
break in the flat-top noise at very low frequencies. 

\citet{WK99} found a correlation between the break frequency and the
frequency of the ``hump'' present in the power spectra of atoll, Z
sources, black hole candidates and AMPs ($L_h$ or sometimes an
associated QPO, $L_{LF}$). We find that the frequency of the first
zero-centered Lorentzian detected above the break in the power spectra
of IGR~J00291+5934 also satisfies this WK correlation
(Figure~\ref{fig:BHWK}, bottom) which leads us to identify it as
$L_h$. The fact that $\nu_{QPO}$ and $\nu_{QPO/2}$ fall much below
this relation (on top of the break as discussed) suggests that they
are not directly related to the low frequency QPOs ($L_{LF}$,
$L_{LF/2}$) measured sometimes on top of the ``hump''. Interestingly,
the same happens for XTE~J1118+480 and XTE~J1550-564: the QPOs on top
of the break fall clearly below the WK correlation, whereas $L_h$
follows it.

\begin{figure}[b]
  \resizebox{1.0\columnwidth}{!}{\rotatebox{-90}{\includegraphics[]{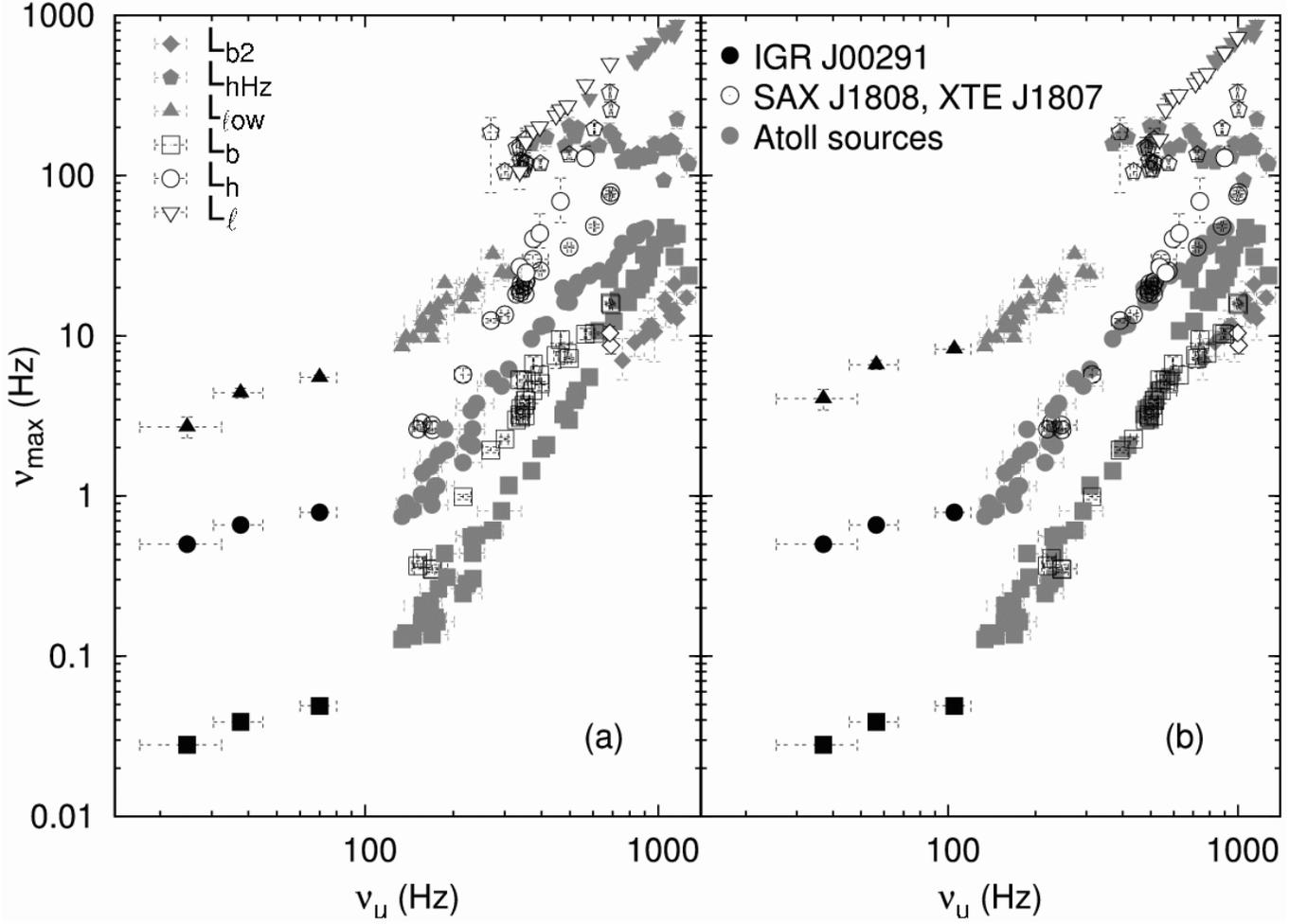}}}

  \caption{Frequency-frequency correlations in neutron star low-mass X-ray binaries. IGR~J00291+5934 (black symbols at $\nu_u \lesssim$100~Hz) is compared with two accreting millisecond pulsars showing shifts in these correlations (SAX~J1808.4--3658 and XTE~J1807--294, open symbols) and with non-pulsing atoll sources (grey symbols, after \citealt{Straaten02}). In a) the characteristic frequencies of power spectral components (identified by the symbols on the left) are plotted versus the frequency of the upper kilohertz QPO (or $L_u$). In b) $\nu_u$ and $\nu_{\ell}$ of SAX~J1808.4--3658 and XTE~J1807--294 have been multiplied by the reported shift factors, whereas a factor of 1.5 has been used for IGR~J00291+5934.}
    \label{fig:vStraaten}
\end{figure}

The relations between the characteristic frequencies of several
variability components of Z and atoll sources and the frequency of the
upper kilohertz QPO, $\nu_u$, were studied and synthesized by
\citet{Straaten03} (Figure~\ref{fig:vStraaten}). Shifts in these
frequency-frequency correlations were found in two AMPs by
\citet{Straaten05} and \citet[][in SAX~J1808.4--3658 and
XTE~J1807--294, respectively]{Linares05}, in a way that was best
explained by the upper and lower kHz QPOs having frequencies ($\nu_u$,
$\nu_{\ell}$) lower than the ones measured in atoll sources by a
factor $\sim$1.5. The component $L_{\ell ow}$ plotted in
Figure~\ref{fig:vStraaten} is the broad component present below $L_u$
(the component with the highest characteristic frequency) in extreme
island states of low-luminosity atoll sources (including bursters and
AMPs; \citealt{Straaten02,Straaten03,Straaten05}). We tentatively
identify the two highest frequency variability components in
IGR~J00291+5934 as $L_u$ and $L_{\ell ow}$, as these are the only
broad variable-frequency components detected in extreme island states
of NS-LMXBs above $L_h$, and plot the frequencies of $L_b$, $L_h$ and
$L_{\ell ow}$ against that of $L_u$ (Figure~\ref{fig:vStraaten}a). As
can clearly be seen in Figure~\ref{fig:vStraaten}, the frequencies of
IGR~J00291+5934 also vary in correlation to one another, but with
slopes much less steep than those found in atoll sources and AMPs at
higher frequencies. In order to compare IGR~J00291+5934 with the two
shifted AMPs, we multiply $\nu_u$ and $\nu_{\ell ow}$ by a trial
factor of 1.5 (Figure~\ref{fig:vStraaten}b). The tracks traced by
IGR~J00291+5934 seem to connect better with the ones at higher
frequencies in the shifted version of the frequency-frequency plots,
but of course the difference in slopes remains. We conclude that a
radical change in the slope of these correlations appears to occur at
frequencies $\nu_u \lesssim$100~Hz. It is unclear at this point to
what extent this is universal to LMXBs or specific to
IGR~J00291+5934. It will be interesting to study the behavior of our
source in future outbursts to see if at $\nu_u \gtrsim$~100Hz its
components follow the usual correlations or not.

\subsection{Mass scaling for break frequencies}
\label{ssec:scaling}

This work also sheds some light on another interesting issue regarding
broad-band noise in power spectra. Break frequencies are often used to
estimate masses of accreting black holes with the argument that
timescales in the accretion disk scale linearly with the mass of the
central object and therefore the observed frequencies should scale
inversely \citep[e.g.,][]{Uttley02}. From the theoretical point of
view this argument is weakened by our poor knowledge of what produces
the broad-band noise and which timescale sets its break frequency
\citep[viscous, dynamical, related with the size of the disk, with
shot lifetime; see
e.g.][]{Priedhorsky79,Manmoto96,Poutanen99,Merloni00,Churazov01}. From
the observational point of view the application of this argument using
break frequencies ($\nu_b$) of the flat-top noise ($L_b$) observed in
low-hard states of LMXBs is clearly challenged by IGR~J00291+5934, as
it harbors a neutron star and shows break frequencies lower than the
ones observed in most of the more massive stellar-mass black
holes. Although $\nu_b$ of Cyg~X-1 in the low-hard state was used in
the past to infer masses of super-massive black holes, recent advances
in active galactic nuclei (AGN) X-ray timing indicate that most AGN
are in a state similar to the high-soft state of BH-LMXBs
\citep{Uttley05}. Any power spectral feature that is used in such
mass-scaling arguments should i) have a rather stable frequency (like
perhaps the hectohertz Lorentzian in NS systems, \citealt{Straaten02};
see also \citealt{Pottschmidt03}) or ii) be accurately corrected for
drifts due to some parameter varying in and between systems (e.g. mass
accretion rate; \citealt{McHardy04,Uttley05}) or a combination of i)
and ii). In any case the power spectra observed in IGR~J00291+5934
refute a simple mass-frequency relation for most of the variability
components present in LMXBs.

\textbf{Acknowledgments:}

We thank Phil Uttley for stimulating discussions on AGN X-ray variability. We also thank D. Altamirano, M. Klein-Wolt, M. M{\' e}ndez and A. Patruno for useful discussions, and the anonymous referee for valuable comments that improved the clarity of this paper.

\clearpage

\begin{table}[t]
\tiny
\caption{Log of the observations.}
\begin{minipage}{\textwidth}
\begin{tabular}{l c c c c c c c r}
\hline\hline
Set & ObsID\footnote{In the first 5 obs. (P90052) GoodXenon modes were used for timing while in the rest (P90425) event modes E\_125us\_64M\_0\_1s were used} & Date\footnote{Observation start date in days since MJD 53341 (December $2^{nd}$, 2004, the discovery date)} & Detectors\footnote{Average number of active detectors} &  Count Rate \footnote{Average and standard deviation of the $\sim$2.5-30 keV count rate, including all active detectors and not corrected for background} & 1024-s PDS\footnote{Number of power density spectra extracted from the observation, each of them 1024 s long} & Total / Bkg. \footnote{Set averages of total and background $\sim$2.5-30 keV count rates (background estimated from set C; see Section~\ref{sec:data})} \\
 & & & & (c/s) & & (c/s) \\
\hline
A1 & 90052-03-01-00 & 0.9 & 3.0 & 308.9$\pm$0.0 & 1 & 232.9 / 61.2 \\
 & 90052-03-01-14 & 2.7 & 3.0 & 262.5$\pm$15.0 & 2 & \\
 & 90052-03-01-04 & 3.6 & 3.0 & 240.4$\pm$4.7 & 12 & \\
 & 90052-03-01-05 & 4.0 & 3.2 & 256.4$\pm$33.7 & 8 & \\
 & 90052-03-01-06 & 4.6 & 4.0 & 300.4$\pm$8.9 & 3  & \\
 & 90425-01-01-08 & 4.9 & 3.3 & 236.5$\pm$29.0 & 7  & \\
 & 90425-01-01-02 & 5.1 & 5.0 & 347.8$\pm$6.3 & 3 & \\
 & 90425-01-01-01 & 5.2 & 5.0 & 351.1$\pm$0.0 & -- & \\
 & 90425-01-01-03 & 5.3 & 5.0 & 369.0$\pm$0.0 & -- & \\
 & 90425-01-01-09 & 5.6 & 3.0 & 209.1$\pm$4.4 & 3  & \\
 & 90425-01-01-07 & 5.9 & 3.5 & 225.5$\pm$31.6 & 12 & \\
 & 90425-01-01-10 & 6.5 & 2.9 & 180.5$\pm$53.6 & 14 & \\
 & 90425-01-01-12 & 6.7 & 3.0 & 188.2$\pm$11.5 & 2 & \\
\hline
A2 & 90425-01-01-110 & 7.0 & 3.3 & 192.6$\pm$41.2 & 13 & 166.6 / 59.2\\
 & 90425-01-01-11 & 7.2 & 3.0 & 166.9$\pm$1.1 & 2 & \\
 & 90425-01-01-15 & 7.3 & 3.0 & 168.3$\pm$2.1 & 2 & \\
 & 90425-01-01-13 & 7.5 & 3.2 & 174.3$\pm$21.3 & 15 & \\
 & 90425-01-02-07 & 7.6 & 3.0 & 156.7$\pm$0.0 & 1 & \\
 & 90425-01-02-08 & 7.7 & 3.0 & 174.6$\pm$15.2 & 2 & \\
 & 90425-01-02-000 & 7.9 & 3.2 & 167.3$\pm$22.2 & 12 & \\
 & 90425-01-02-00 & 8.1 & 3.0 & 148.2$\pm$2.4 & 4 & \\
 & 90425-01-02-15 & 8.2 & 3.0 & 150.3$\pm$2.4 & 2 & \\
 & 90425-01-02-04 & 8.3 & 3.0 & 144.6$\pm$3.8 & 2 & \\
 & 90425-01-02-05 & 8.4 & 3.0 & 142.6$\pm$2.6 & 3 & \\
 & 90425-01-02-16 & 8.5 & 3.0 & 141.5$\pm$4.5 & 10 & \\
\hline
B & 90425-01-02-02 & 8.7 & 3.0 & 127.9$\pm$0.0 & 1 & 94.1 / 53.9\\
 & 90425-01-02-030 & 8.9 & 3.0 & 126.6$\pm$5.4 & 12 & \\
 & 90425-01-02-03 & 9.1 & 3.0 & 117.5$\pm$3.7 & 5 & \\
 & 90425-01-02-09 & 9.2 & 3.0 & 112.6$\pm$3.5 & 2 & \\
 & 90425-01-02-100 & 9.4 & 2.8 & 102.3$\pm$14.0 & 15 & \\
 & 90425-01-02-10 & 9.6 & 3.0 & 126.0$\pm$0.0 & 1 & \\
 & 90425-01-02-11 & 9.7 & 3.0 & 99.9$\pm$0.0 & 1 & \\
 & 90425-01-02-120 & 9.9 & 2.8 & 91.8$\pm$9.2 & 12 & \\
 & 90425-01-02-12 & 10.1 & 2.8 & 81.9$\pm$10.7 & 5 & \\
 & 90425-01-02-13 & 10.3 & 3.0 & 85.9$\pm$2.7 & 3 & \\
 & 90425-01-02-14 & 10.4 & 2.6 & 77.7$\pm$13.9 & 11& \\
 & 90425-01-02-17 & 10.6 & 2.0 & 65.7$\pm$7.2 & 2 & \\
 & 90425-01-02-06 & 11.5 & 3.0 & 72.5$\pm$3.8 & 5 & \\
 & 90425-01-02-01 & 11.6 & 2.0 & 52.1$\pm$0.0 & 1 & \\
 & 90425-01-02-18 & 11.9 & 3.3 & 74.0$\pm$17.6 & 10 & \\
 & 90425-01-02-20 & 13.2 & 3.0 & 58.1$\pm$0.0 & 1 & \\
\hline
C\footnote{Set C contains mainly background/foreground observations.} & 90425-01-02-25 & 14.1 & 4.0 & 72.4$\pm$0.4 & 2 & 63.6 / 63.6\\
 & 90425-01-02-26 & 14.4 & 2.0 & 40.0$\pm$0.0 & 1 & \\
 & 90425-01-02-24 & 14.5 & 4.0 & 77.4 $\pm$3.7 & 3 & \\
 & 90425-01-02-19 & 14.5 & 3.0 & 72.8$\pm$0.0 & 1 & \\
 & 90425-01-02-27 & 14.6 & 5.0 & 91.5$\pm$0.0 & 1 & \\
 & 90425-01-03-00 & 14.9 & 3.3 & 63.5$\pm$8.1 & 9 & \\
 & 90425-01-03-01 & 15.3 & 3.0 & 56.1$\pm$2.5 & 8 & \\
 & 90425-01-03-02 & 16.0 & 4.0 & 72.7$\pm$1.7 & 2 & \\
 & 90425-01-03-03 & 16.2 & 3.0 & 55.2$\pm$1.5 & 6 & \\
 & 90425-01-03-04 & 17.3 & 3.4 & 66.0$\pm$10.1 & 7 & \\
 & 90425-01-03-05 & 17.9 & 4.0 & 71.9$\pm$2.1 & 6 & \\
 & 90425-01-03-06 & 18.2 & 3.0 & 54.8$\pm$2.0 & 5 & \\
\hline\hline
\end{tabular}
\end{minipage}
\label{table:obs}
\end{table}

\clearpage

\begin{table}[t]
\footnotesize
\caption{Best fit parameters from 6-Lorentzian model and integrated rms variability}
\begin{minipage}{\textwidth}
\begin{tabular}{l l c c c c c c r r}
\hline\hline
Set & Parameter & $L_b$ & $L_{QPO/2}$ & $L_{QPO}$ & $L_h$ & $L_{\ell ow}$ & $L_u$ & $\chi^2$/d.o.f. & rms (\%) \\
\hline
A1 & $\nu_{max}(Hz)$ & (4.9$\pm$0.2)$\times10^{-2}$& (2.15$\pm$0.05)$\times10^{-2}$ & (4.43$\pm$0.10)$\times10^{-2}$ & 0.79$\pm$0.03 & 5.5$\pm$0.2 & 70$\pm$10 & 357/325 & 41.7$\pm$0.4 \\
 & rms (\%) & 24.6$\pm$0.3 & 4.8$\pm$0.8 & 4.5$\pm$0.7 & 23.4$\pm$0.2 & 22.1$\pm$0.2 & 15.5$\pm$0.4 & \\
 & Q & (4$\pm$4)$\times10^{-2}$ & 5.5$\pm$3.2 & 6.2$\pm$2.2 & 0 (fixed) & 0 (fixed) & 0 (fixed) & \\
\hline
A2 & $\nu_{max}(Hz)$ & (3.9$\pm$0.1)$\times10^{-2}$ & (2.11$\pm$0.09)$\times10^{-2}$ & (4.03$\pm$0.06)$\times10^{-2}$ &  0.66$\pm$0.02 & 4.4$\pm$0.3 & 37.6$\pm$7.2 & 335/326  & 47.3$\pm$0.6\\
 & rms (\%) & 29.3$\pm$0.3 & 5.5$\pm$1.0 & 4.4$\pm$0.6 & 27.1$\pm$0.3 & 23.5$\pm$0.4 & 15.6 $\pm$ 0.7 & \\
 & Q & 0 (fixed) & 5.1$\pm$2.5 & 16$\pm$6 & 0 (fixed) & 0 (fixed) & 0 (fixed) & \\
\hline
B & $\nu_{max}(Hz)$ & (2.8$\pm$0.1)$\times10^{-2}$ & -- & -- & 0.50$\pm$0.03 & 2.7$\pm$0.4 & 25.0$\pm$7.7 & 418/332  & 57.8$\pm$1.2 \\
 & rms (\%) & 37.5$\pm$0.3 & -- & -- & 34.4$\pm$0.4 & 25.9$\pm$1.0 & 21.2$\pm$1.4 & \\
 & Q & 0 (fixed) & -- & -- &  0 (fixed)  & 0 (fixed) & 0 (fixed) & \\
\hline\hline
\end{tabular}
\end{minipage}
\label{table:psfits}
\end{table}


\begin{table}[t]
\footnotesize
\caption{Best fit parameters from 7-Lorentzian model}
\begin{minipage}{\textwidth}
\begin{tabular}{l l c c c c c c c r}
\hline\hline
Set & Parameter & $L_{1b}$ & $L_{QPO/2}$ & $L_{QPO}$ & $L_{2b}$ & $L_h$ & $L_{\ell ow}$ & $L_u$ & $\chi^2$/d.o.f.\\
\hline
A1 & $\nu_{max}(Hz)$ & (1.6$\pm$0.3)$\times10^{-2}$& (2.25$\pm$0.05)$\times10^{-2}$ & (4.36$\pm$0.08)$\times10^{-2}$ & (7.8$\pm$0.8)$\times10^{-2}$ & 0.71$\pm$0.03 & 5.3$\pm$0.2 & 66.3$\pm$9.3 & 327/322\\
 & rms (\%) & 12.8$\pm$2.1 & 6.5$\pm$2.1 (2.3$\sigma$) & 8.3$\pm$1.6 & 17.8$\pm$1.7 & 24.0$\pm$0.2 & 22.3$\pm$0.2 & 15.6$\pm$0.4 \\
 & Q & 0.53$\pm$0.17 & 3.6$\pm$1.6 & 2.9$\pm$1.0 & 0.44$\pm$0.11 & 0 (fixed) & 0 (fixed) & 0 (fixed)\\
\hline
A2 & $\nu_{max}(Hz)$ & (1.8$\pm$0.4)$\times10^{-2}$ & (2.19$\pm$0.08)$\times10^{-2}$ & (3.98$\pm$0.05)$\times10^{-2}$ & (6.5$\pm$0.9)$\times10^{-2}$ &  0.63$\pm$0.03 & 4.3$\pm$0.3 & 36.7$\pm$6.9 & 323/322 \\
 & rms (\%) & 18.7$\pm$3.4 & 5.8$\pm$2.0 (1.8$\sigma$) & 6.1$\pm$0.9 & 21.0$\pm$3.1 & 27.6$\pm$0.3 & 23.7$\pm$0.4 & 15.7$\pm$0.7\\
 & Q & 0.3$\pm$0.1 & $5^{+9}_{-2}$ & 7.8$\pm$2.5 & 0.3$\pm$0.1 & 0 (fixed) & 0 (fixed) & 0 (fixed) \\
\hline
B & $\nu_{max}(Hz)$ & (1.8$\pm$0.3)$\times10^{-2}$ & -- & -- & (5.0$\pm$1.3)$\times10^{-2}$ & 0.56$\pm$0.03 & 3.1$\pm$0.5 & $27^{+12}_{-7}$ & 403/330 \\
 & rms (\%) & 29.5$\pm$4.1 & -- & -- & 25.2$\pm$4.7 & 34.5$\pm$0.7 & 24.9$\pm$1.1 & 20.7$\pm$1.5\\
 & Q & 0 (fixed) & -- & -- & 0 (fixed) & 0 (fixed)  & 0 (fixed) & 0 (fixed)\\
\hline\hline
\footnotetext{Within this function two Lorentzians are used to fit the break in the flat-top noise, $L_{1b}$ and $L_{2b}$ (see Sec.~\ref{ssec:timing})}
\end{tabular}
\end{minipage}
\label{table:psfits2}
\end{table}


\begin{thebibliography}{68}
\expandafter\ifx\csname natexlab\endcsname\relax\def\natexlab#1{#1}\fi
\expandafter\ifx\csname url\endcsname\relax
  \def\url#1{{\tt #1}}\fi
\expandafter\ifx\csname urlprefix\endcsname\relax\def\urlprefix{URL }\fi

\bibitem[{{Altamirano} et~al.(2005){Altamirano}, {van der Klis}, {Mendez}
  et~al.}]{Altamirano05}
{Altamirano} D., {van der Klis} M., {Mendez} M., et~al., 2005, \apj, 633, 358

\bibitem[{{Barret} et~al.(2000){Barret}, {Olive}, {Boirin} et~al.}]{Barret00}
{Barret} D., {Olive} J.F., {Boirin} L., et~al., 2000, \apj, 533, 329

\bibitem[{{Belloni} \& {Hasinger}(1990)}]{BH90}
{Belloni} T., {Hasinger} G., 1990, \aap, 227, L33

\bibitem[{{Belloni} et~al.(2002){Belloni}, {Psaltis}, \& {van der
  Klis}}]{Belloni02}
{Belloni} T., {Psaltis} D., {van der Klis} M., 2002, \apj, 572, 392

\bibitem[{{Bhattacharya} \& {van den Heuvel}(1991)}]{BhatHeuv91}
{Bhattacharya} D., {van den Heuvel} E.P.J., 1991, \physrep, 203, 1

\bibitem[{{Burderi} et~al.(2005){Burderi}, {Di Salvo}, {Riggio}
  et~al.}]{Burderi05}
{Burderi} L., {Di Salvo} T., {Riggio} A., et~al., 2005, astro-ph/0509224

\bibitem[{{Casella} et~al.(2005){Casella}, {Belloni}, \& {Stella}}]{Casella05}
{Casella} P., {Belloni} T., {Stella} L., 2005, \apj, 629, 403

\bibitem[{{Chen} \& {Taam}(1994)}]{Chen94}
{Chen} X., {Taam} R.E., 1994, \apj, 431, 732

\bibitem[{{Churazov} et~al.(2001){Churazov}, {Gilfanov}, \&
  {Revnivtsev}}]{Churazov01}
{Churazov} E., {Gilfanov} M., {Revnivtsev} M., 2001, \mnras, 321, 759

\bibitem[{{Cui} et~al.(1999){Cui}, {Zhang}, {Chen}, \& {Morgan}}]{Cui99}
{Cui} W., {Zhang} S.N., {Chen} W., {Morgan} E.H., 1999, \apjl, 512, L43

\bibitem[{{Cumming} et~al.(2001){Cumming}, {Zweibel}, \&
  {Bildsten}}]{Cumming01}
{Cumming} A., {Zweibel} E., {Bildsten} L., 2001, \apj, 557, 958

\bibitem[{{de Martino} et~al.(2004){de Martino}, {Matt}, {Belloni}, {Haberl},
  \& {Mukai}}]{Martino04}
{de Martino} D., {Matt} G., {Belloni} T., {Haberl} F., {Mukai} K., 2004, \aap,
  415, 1009

\bibitem[{{di Salvo} et~al.(2004){di Salvo}, {Santangelo}, \&
  {Segreto}}]{DiSalvo04}
{di Salvo} T., {Santangelo} A., {Segreto} A., 2004, Nuclear Physics B
  Proceedings Supplements, 132, 446

\bibitem[{{Eckert} et~al.(2004){Eckert}, {Walter}, {Kretschmar}
  et~al.}]{Eckert04}
{Eckert} D., {Walter} R., {Kretschmar} P., et~al., 2004, The Astronomer's
  Telegram, 352, 1

\bibitem[{{Falanga} et~al.(2005{\natexlab{a}}){Falanga}, {Bonnet-Bidaud},
  {Poutanen} et~al.}]{Falanga05}
{Falanga} M., {Bonnet-Bidaud} J.M., {Poutanen} J., et~al., 2005{\natexlab{a}},
  astro-ph/0503292

\bibitem[{{Falanga} et~al.(2005{\natexlab{b}}){Falanga}, {Bonnet-Bidaud}, \&
  {Suleimanov}}]{Falanga05c}
{Falanga} M., {Bonnet-Bidaud} J.M., {Suleimanov} V., 2005{\natexlab{b}}, \aap,
  444, 561

\bibitem[{{Falanga} et~al.(2005{\natexlab{c}}){Falanga}, {Kuiper}, {Poutanen}
  et~al.}]{Falanga05b}
{Falanga} M., {Kuiper} L., {Poutanen} J., et~al., 2005{\natexlab{c}}, \aap,
  444, 15

\bibitem[{{Fender} et~al.(2004){Fender}, {De Bruyn}, {Pooley}, \&
  {Stappers}}]{Fender04}
{Fender} R., {De Bruyn} G., {Pooley} G., {Stappers} B., 2004, The Astronomer's
  Telegram, 361, 1

\bibitem[{{Fox} \& {Kulkarni}(2004)}]{Fox04}
{Fox} D.B., {Kulkarni} S.R., 2004, The Astronomer's Telegram, 354, 1

\bibitem[{{Galloway} et~al.(2005){Galloway}, {Markwardt}, {Morgan},
  {Chakrabarty}, \& {Strohmayer}}]{Galloway05}
{Galloway} D.K., {Markwardt} C.B., {Morgan} E.H., {Chakrabarty} D.,
  {Strohmayer} T.E., 2005, \apjl, 622, L45

\bibitem[{{Garcia} et~al.(2001){Garcia}, {McClintock}, {Narayan}
  et~al.}]{Garcia01}
{Garcia} M.R., {McClintock} J.E., {Narayan} R., et~al., 2001, \apjl, 553, L47

\bibitem[{{Ghosh} \& {Lamb}(1978)}]{Gosh78}
{Ghosh} P., {Lamb} F.K., 1978, \apjl, 223, L83

\bibitem[{{Gilfanov} et~al.(2003){Gilfanov}, {Revnivtsev}, \&
  {Molkov}}]{Gilfanov03}
{Gilfanov} M., {Revnivtsev} M., {Molkov}, S., 2003, \aap,
  410, 217

\bibitem[{{Haberl} \& {Motch}(1995)}]{Haberl95}
{Haberl} F., {Motch} C., 1995, \aap, 297, L37+

\bibitem[{{Hasinger} \& {van der Klis}(1989)}]{Hasinger89}
{Hasinger} G., {van der Klis} M., 1989, \aap, 225, 79

\bibitem[{{Jonker} et~al.(2005){Jonker}, {Campana}, {Steeghs}
  et~al.}]{Jonker05}
{Jonker} P.G., {Campana} S., {Steeghs} D., et~al., 2005, \mnras, 361, 511

\bibitem[{{Klein-Wolt}(2004)}]{Kleinwolt04}
{Klein-Wolt} M., 2004, Thesis, Universiteit van Amsterdam

\bibitem[{{Klein-Wolt} \& {van der Klis}(2006)}]{Kleinwolt06}
{Klein-Wolt} M., {van der Klis} M., 2006, in prep.

\bibitem[{{Klein-Wolt} et~al.(2004){Klein-Wolt}, {Homan}, \& {van der
  Klis}}]{Kleinwolt04b}
{Klein-Wolt} M., {Homan} J., {van der Klis} M., 2004, Nuclear Physics B
  Proceedings Supplements, 132, 381

\bibitem[{{Kuulkers} et~al.(1994){Kuulkers}, {van der Klis}, {Oosterbroek}
  et~al.}]{Kuulkers94}
{Kuulkers} E., {van der Klis} M., {Oosterbroek} T., et~al., 1994, \aap, 289,
  795

\bibitem[{{Lamb} et~al.(1973){Lamb}, {Pethick}, \& {Pines}}]{Lamb73}
{Lamb} F.K., {Pethick} C.J., {Pines} D., 1973, \apj, 184, 271

\bibitem[{{Linares} et~al.(2005){Linares}, {van der Klis}, {Altamirano}, \&
  {Markwardt}}]{Linares05}
{Linares} M., {van der Klis} M., {Altamirano} D., {Markwardt} C.B., 2005, \apj,
  634, 1250

\bibitem[{{Manmoto} et~al.(1996){Manmoto}, {Takeuchi}, {Mineshige},
  {Matsumoto}, \& {Negoro}}]{Manmoto96}
{Manmoto} T., {Takeuchi} M., {Mineshige} S., {Matsumoto} R., {Negoro} H., 1996,
  \apjl, 464, L135+

\bibitem[{{Markwardt} et~al.(2004{\natexlab{a}}){Markwardt}, {Galloway},
  {Chakrabarty}, {Morgan}, \& {Strohmayer}}]{Markwardt04c}
{Markwardt} C.B., {Galloway} D.K., {Chakrabarty} D., {Morgan} E.H.,
  {Strohmayer} T.E., 2004{\natexlab{a}}, The Astronomer's Telegram, 360, 1

\bibitem[{{Markwardt} et~al.(2004{\natexlab{b}}){Markwardt}, {Swank}, \&
  {Strohmayer}}]{Markwardt04b}
{Markwardt} C.B., {Swank} J.H., {Strohmayer} T.E., 2004{\natexlab{b}}, The
  Astronomer's Telegram, 353, 1

\bibitem[{{McHardy} et~al.(2004){McHardy}, {Papadakis}, {Uttley}, {Page}, \&
  {Mason}}]{McHardy04}
{McHardy} I.M., {Papadakis} I.E., {Uttley} P., {Page} M.J., {Mason} K.O., 2004,
  \mnras, 348, 783

\bibitem[{{Merloni} et~al.(2000){Merloni}, {Di Matteo}, \&
  {Fabian}}]{Merloni00}
{Merloni} A., {Di Matteo} T., {Fabian} A.C., 2000, \mnras, 318, L15

\bibitem[{{Migliari} \& {Fender}(2006)}]{Migliari06}
{Migliari} S., {Fender} R.P., 2006, \mnras, 366, 79

\bibitem[{{Norton} et~al.(1999){Norton}, {Beardmore}, {Allan}, \&
  {Hellier}}]{Norton99}
{Norton} A.J., {Beardmore} A.P., {Allan} A., {Hellier} C., 1999, \aap, 347, 203

\bibitem[{{Olive} et~al.(1998){Olive}, {Barret}, {Boirin} et~al.}]{Olive98}
{Olive} J.F., {Barret} D., {Boirin} L., et~al., 1998, \aap, 333, 942

\bibitem[{{Patterson}(1994)}]{Patterson}
{Patterson} J., Mar. 1994, \pasp, 106, 209

\bibitem[{{Pooley}(2004)}]{Pooley04}
{Pooley} G., 2004, The Astronomer's Telegram, 355, 1

\bibitem[{{Pottschmidt} et~al.(2003){Pottschmidt}, {Wilms}, {Nowak}
  et~al.}]{Pottschmidt03}
{Pottschmidt} K., {Wilms} J., {Nowak} M.A., et~al., 2003, \aap, 407, 1039

\bibitem[{{Poutanen} \& {Fabian}(1999)}]{Poutanen99}
{Poutanen} J., {Fabian} A.C., 1999, \mnras, 306, L31

\bibitem[{{Priedhorsky} et~al.(1979){Priedhorsky}, {Garmire}, {Rothschild}
  et~al.}]{Priedhorsky79}
{Priedhorsky} W., {Garmire} G.P., {Rothschild} R., et~al., 1979, \apj, 233, 350

\bibitem[{{Psaltis} \& {Chakrabarty}(1999)}]{Psaltis99c}
{Psaltis} D., {Chakrabarty} D., 1999, \apj, 521, 332

\bibitem[{{Rappaport} et~al.(2004){Rappaport}, {Fregeau}, \&
  {Spruit}}]{Rappaport04}
{Rappaport} S.A., {Fregeau} J.M., {Spruit} H., 2004, \apj, 606, 436

\bibitem[{{Remillard}(2004)}]{Remillard04}
{Remillard} R., 2004, The Astronomer's Telegram, 357, 1

\bibitem[{{Revnivtsev} et~al.(2000){Revnivtsev}, {Sunyaev}, \&
  {Borozdin}}]{Revni00}
{Revnivtsev} M., {Sunyaev} R., {Borozdin} K., 2000, \aap, 361, L37

\bibitem[{{Rupen} et~al.(2004){Rupen}, {Dhawan}, \& {Mioduszewski}}]{Rupen04}
{Rupen} M.P., {Dhawan} V., {Mioduszewski} A.J., 2004, The Astronomer's
  Telegram, 364, 1

\bibitem[{{Shaw} et~al.(2005){Shaw}, {Mowlavi}, {Rodriguez} et~al.}]{Shaw05}
{Shaw} S.E., {Mowlavi} N., {Rodriguez} J., et~al., 2005, \aap, 432, L13

\bibitem[{{Steeghs} et~al.(2004){Steeghs}, {Blake}, {Bloom} et~al.}]{Steeghs04}
{Steeghs} D., {Blake} C., {Bloom} J.S., et~al., 2004, The Astronomer's
  Telegram, 363, 1

\bibitem[{{Sunyaev} \& {Revnivtsev}(2000)}]{SunRev00}
{Sunyaev} R., {Revnivtsev} M., 2000, \aap, 358, 617

\bibitem[{{Sunyaev} et~al.(1991){Sunyaev}, {Arefev}, {Borozdin}
  et~al.}]{Sunyaev91}
{Sunyaev} R.A., {Arefev} V.A., {Borozdin} K.N., et~al., 1991, Soviet Astronomy
  Letters, 17, 409

\bibitem[{{Titarchuk} et~al.(2002){Titarchuk}, {Cui}, \& {Wood}}]{Titarchuk02}
{Titarchuk} L., {Cui} W., {Wood} K., 2002, \apjl, 576, L49

\bibitem[{{Uttley} et~al.(2002){Uttley}, {McHardy}, \& {Papadakis}}]{Uttley02}
{Uttley} P., {McHardy} I.M., {Papadakis} I.E., 2002, \mnras, 332, 231

\bibitem[{{Uttley} \& {McHardy}(2005)}]{Uttley05}
{Uttley} P., {McHardy} I.M., 2005, \mnras, 363, 586

\bibitem[{{van der Klis}(1994)}]{Klis94}
{van der Klis} M., 1994, \apjs, 92, 511

\bibitem[{{van der Klis}(1995)}]{vanderklis95b}
{van der Klis} M., 1995, Proceedings of the NATO Advanced Study Institute on
  the Lives of the Neutron Stars, held in Kemer, Turkey, August 19-September
  12, 1993. Editor(s), M. A. Alpar, U. Kiziloglu, J. van Paradijs; Publisher,
  Kluwer Academic, Dordrecht, The Netherlands, Boston, Massachusetts, 301

\bibitem[{{van der Klis}(2000)}]{vanderKlis00}
{van der Klis} M., 2000, \araa, 38, 717

\bibitem[{{van der Klis}(2006)}]{vanderKlis06}
{van der Klis} M., 2006, in "Compact Stellar X-ray Sources", ed. W. H. G. Lewin
  \& M. van der Klis (Cambridge Univ. Press) (astro-ph/0410551), 39--112

\bibitem[{{van Straaten} et~al.(2002){van Straaten}, {van der Klis}, {di
  Salvo}, \& {Belloni}}]{Straaten02}
{van Straaten} S., {van der Klis} M., {di Salvo} T., {Belloni} T., 2002, \apj,
  568, 912

\bibitem[{{van Straaten} et~al.(2003){van Straaten}, {van der Klis}, \& {M{\'
  e}ndez}}]{Straaten03}
{van Straaten} S., {van der Klis} M., {M{\' e}ndez} M., 2003, \apj, 596, 1155

\bibitem[{{van Straaten} et~al.(2005){van Straaten}, {van der Klis}, \&
  {Wijnands}}]{Straaten05}
{van Straaten} S., {van der Klis} M., {Wijnands} R., 2005, \apj, 619, 455

\bibitem[{{Vikhlinin} et~al.(1994){Vikhlinin}, {Churazov}, \&
  {Gilfanov}}]{Vikh94}
{Vikhlinin} A., {Churazov} E., {Gilfanov} M., 1994, \aap, 287, 73

\bibitem[{{Wijnands}(2005)}]{Wijnands05}
{Wijnands} R., 2005, astro-ph/0501264

\bibitem[{{Wijnands} \& {van der Klis}(1998)}]{Wijnands98}
{Wijnands} R., {van der Klis} M., 1998, \nat, 394, 344

\bibitem[{{Wijnands} \& {van der Klis}(1999)}]{WK99}
{Wijnands} R., {van der Klis} M., 1999, \apj, 514, 939

\bibitem[{{Zhang} et~al.(1995){Zhang}, {Jahoda}, {Swank}, {Morgan}, \&
  {Giles}}]{Zhang95}
{Zhang} W., {Jahoda} K., {Swank} J.H., {Morgan} E.H., {Giles} A.B., 1995, \apj,
  449, 930

\end{thebibliography}
\end{document}